\documentclass[aps,prb,twocolumn,superscriptaddress,floatfix,longbibliography]{revtex4-2}

\usepackage{amsmath}
\usepackage{graphicx}
\usepackage{dcolumn}
\usepackage{bm}
\usepackage{xcolor}
\usepackage{physics}

\begin{document}
\title{Mixed-State Berry Curvature in quantum multiparameter estimations}
\author{Xiaoguang Wang}
\affiliation{Zhejiang Key Laboratory of Quantum State Control and Optical Field Manipulation, Department of Physics, Zhejiang Sci-Tech University, 310018 Hangzhou, China}
\email{xgwang@zstu.edu.cn}
\author{Xiao-Ming Lu}
\affiliation{School of Sciences, Hangzhou Dianzi University, Hangzhou 310018, China}
\email{lxm@hdu.edu.cn}
\author{Yunbo Zhang}
\affiliation{Zhejiang Key Laboratory of Quantum State Control and Optical Field Manipulation, Department of Physics, Zhejiang Sci-Tech University, 310018 Hangzhou, China}
\author{Libin Fu}
\affiliation{Graduate School of China Academy of Engineering Physics, Beijing 100193, China}
\author{Shu Chen}
\affiliation{Beijing National Laboratory for Condensed Matter Physics, Institute of Physics, Chinese Academy of Sciences, Beijing 100190, China}
\date{\today}

\begin{abstract}
	For pure states, the quantum Berry curvature was well studied. However, the quantum curvature for mixed states has received less attention. From the concept of symmetric logarithmic derivative, we introduce a mixed-state quantum curvature and find that it plays a key role in the field of multi-parameter precision estimations. Through spectral decomposition, we derive the mixed-state Berry curvature for both the full-rank and non-full-rank density matrices. As an example, we obtain the exact expression of the Berry curvature for an arbitrary qubit state.
\end{abstract}
\maketitle

{\em Introduction.---}One of central objects in the field of quantum geometry is the Berry curvature~\cite{Berry1984} and the integration of the Berry curvature over the Brillouin zone leads to the Chern number to characterize topological phases.  Berry curvature can induce the spin Hall effect~\cite{Son2012,SCZhang} and can be considered as a kind of magnetic field in momentum space, leading to anomalous Hall effect~\cite{AHall} and anomalous Nernst effect~\cite{Nernst}.

In single-flavor color superconductor systems, the interplay between Berry curvature and topological properties was investigated recently~\cite{bcsc}.
By reconstructing Berry curvature through Hall drift,
Chern number can be measured experimentally~\cite{SandersPan} and 
through resonant infrared
magnetic circular dichroism, the Berry curvature was probed in magnetic topological insulators~\cite{bcexp1}. Gritsev and
Polkovnikov  used the result from adiabatic perturbation theory to study a slowly
driven system, and the Berry curvature emerges due to a certain 
quench~\cite{gritsev2012dynamical}.

Geometric phases for pure states was generalized to the case of mixed states~\cite{sjoqvist2000geometric}. However,  much less attention was devoted to the Berry curvature for mixed states. 
In the field of quantum multiparameter estimation, a trade-off relation was derived between the regrets of Fisher information about different parameters. 
The inequality is given by~\cite{luwang}
\begin{equation}\label{trade}
	\Delta_\alpha^2+\Delta_\beta^2+2\sqrt{1-C_{\alpha\beta}^2}\Delta_\alpha\Delta_\beta\ge C_{\alpha\beta}^2,
\end{equation}
where the regret for parameter $\alpha$ is
\begin{equation}
	\Delta_\alpha^2=({\cal F}_\alpha-F_\alpha)/{\cal F}_\alpha
\end{equation}
and 
\begin{align}
	C_{\alpha\beta}^2=|{\rm Im}\langle L_\alpha L_\beta\rangle|^2/({\cal F}_\alpha {\cal F}_\beta) 
\end{align}
with the notation \(\ev{X}=\tr(\rho X)\) denoting the quantum expectation value of any operator \(X\) on the state \(\rho\).
The quantities ${\cal F}_\alpha$ is the quantum Fisher information (QFI) and $F_\alpha$ is the classical Fisher information with respect to a special quantum measurement. 
The Hermitian operator $L_\alpha$ is the symmetric logarithmic derivative (SLD) operator with respect to parameter $\alpha$ and for a density operator $\rho$ depending on the parameter $\alpha$, it
is defined as
\begin{equation}\label{sld}
	\partial_\alpha\rho=\frac{1}2(L_\alpha\rho+\rho L_\alpha).
\end{equation}
No matter the density operator is pure or mixed, the QFI is given by a uniform expression~\cite{helstrom1969,holevo2011probabilistic}
\begin{align}\label{qfi}
	{\cal F}_\alpha=\text{Tr}(\rho L_\alpha^2).
\end{align}
Since the expectation value of the SLD operator on the state $\rho$ is zero, the QFI is the variance of the SLD operator. It should be noted that practical schemes were given to measure the QFI~\cite{xingyu,wenkuiprb}. In addition, the trade-off relation given by the inequality was further investigated from experiments~\cite{conlon2023approaching,xia2023toward}.

We observe that except for the two Fisher information in inequality \eqref{trade}, there appears a key quantity ${\rm Im}\langle L_\alpha L_\beta\rangle$, which can be written as
\begin{equation}
	{\rm Im}\langle L_\alpha L_\beta\rangle=\frac{1}2{\rm Im}\langle [L_\alpha, L_\beta]\rangle=-\frac{i}2\langle [L_\alpha, L_\beta]\rangle.
\end{equation}
Actually, considering two SLDs, $L_\alpha$ and $L_\beta$, we have the Heisenberg
uncertainty relation
\begin{equation}
	\langle (\Delta L_\alpha)^2 \rangle \langle (\Delta L_\beta)^2 \rangle\geq \frac{|\langle[L_\alpha,L_\beta]\rangle|^2}{4} .\label{HeisenbergUncertaintyRelation12}
\end{equation}
In terms of the QFI given by Eq.~\eqref{qfi}, the above inequality can be rewritten as
\begin{equation}\label{trade222}
	{\cal F}_\alpha {\cal F}_\beta \geq |{\rm Im}\langle L_\alpha L_\beta\rangle|^2=({\rm Im}\langle L_\alpha L_\beta\rangle)^2 .
\end{equation}
We see that the multiplication of two Fisher information is bounded below by the square of the quantity ${\rm Im}\langle L_\alpha L_\beta\rangle$. 

We note that reduction of the regrets
of Fisher information about different parameters is
restricted by a nonzero value of ${\rm Im}\langle L_\alpha L_\beta\rangle$. 
When ${\rm Im}\langle L_\alpha L_\beta\rangle=0$, it is easy to see that inequality
\eqref{trade} becomes trivial, i.e., there is no restrictions on the two regrets.
Similarly, from Eq.~\eqref{trade222}, there is a restriction on the two QFIs only when ${\rm Im}\langle L_\alpha L_\beta\rangle$ is not zero. Therefore, we conclude that the quantity plays a key role in the field of quantum multi-parameter estimation and it is just the quantum curvature given below up to a multiplicative constant.

{\em Quantum curvature for mixed states.---} For a state $|\psi\rangle$ with two parameters $\alpha$ and $\beta$, the Berry curvature is defined as
\begin{align}\label{bc}
	{\Omega}_{\alpha\beta}(\psi)&=i(\langle\partial_{\alpha}\psi|\partial_{\beta}\psi\rangle-\langle\partial_{\beta}\psi|\partial_{\alpha}\psi\rangle)\notag\\
	&=-2{\rm Im}(\langle\partial_{\alpha}\psi|\partial_{\beta}\psi\rangle),
\end{align}
We now try to give a new form of the Berry curvature for pure states in terms of SLDs. Let us consider a quantum pure state written in the form of a density matrix $\rho=|\psi\rangle\langle\psi|$. It follows $\rho^{2}=\rho$ that
\begin{equation}
	\partial_{\alpha}\rho=\partial_{\alpha}\rho^{2}=\left(\partial_{\alpha}\rho\right)\rho+\rho\left(\partial_{\alpha}\rho\right),
\end{equation}
implying that SLD operator can be written as
\begin{equation}\label{sld1}
	L_{\alpha}=2\partial_{\alpha}\rho,
\end{equation}
where we have used Eq.~\eqref{sld}. Furthermore, the SLDs for parameters $\alpha,\beta $ are given by
\begin{align}
	L_{\alpha}&=2(|\partial_{\alpha}\psi\rangle\langle\psi|+|\psi\rangle\langle\partial_{\alpha}\psi|),\\
	L_{\beta}&=2(|\partial_{\beta}\psi\rangle\langle\psi|+|\psi\rangle\langle\partial_{\beta}\psi|),
\end{align}	
From these two SLDs, a straightforward calculation leads to
\begin{equation}\label{llalphabeta}
	\begin{aligned}
		\langle \psi| L_{\alpha}L_{\beta}|\psi\rangle =4(\langle\psi|\partial_{\beta}\psi\rangle\langle\psi|\partial_{\alpha}\psi\rangle
		+\langle\partial_{\alpha}\psi|\partial_{\beta}\psi\rangle),\\
		\langle \psi| L_{\beta}L_{\alpha}|\psi\rangle =4(\langle\psi|\partial_{\alpha}\psi\rangle\langle\psi|\partial_{\beta}\psi\rangle
		+\langle\partial_{\beta}\psi|\partial_{\alpha}\psi\rangle).
	\end{aligned}
\end{equation}
Combining the above equation and Eq.~\eqref{bc}, one finds
\begin{equation}
	{\Omega_{\alpha\beta}(\psi)=\frac{i}{4}\langle\psi|[L_{\alpha},L_{\beta}]|\psi\rangle.}
\end{equation}
The Berry curvature is written as a expectation value of the commutator of two SLDs.

It is natural to define the quantum curvature  for an mixed state $\rho$ as
\begin{align}\label{bcmix}
	{\Omega}_{\alpha\beta}(\rho)=\frac{i}{4}{\rm Tr}(\rho[L_{\alpha},L_{\beta}])=-\frac{1}2{\rm Im}\langle L_\alpha L_\beta\rangle,
\end{align}	
which differs from the quantity ${\rm Im}\langle L_\alpha L_\beta\rangle$ by a multiplicative factor $-1/2$. This quantum curvature directly reduces to the Berry curvature when the mixed state becomes pure. Here the two SLDs was defined for any state $\rho$ with two parameters. Thus, it is
general and applicable for any quantum systems. Note that the quantum curvature is the expectation value of Hermitian operator $i[L_\alpha,L_\beta]/4$. Similarly, the elements of the 
Fisher information matrix 
\begin{equation}
{\cal F}_{\alpha\beta}=\langle[L_\alpha,L_\beta]_+/2\rangle
\end{equation}
is the expectation of anticommutator $[L_\alpha,L_\beta]_+/2$. 

Carollo \textit{et al.}~called the quantum curvature as the mean Uhlmann curvature defined as the Uhlmann geometric phase~\cite{uhlmann1986parallel,carollo2019quantumness} per unit area of a density matrix evolving along an infinitesimal loop in the parameter space and the curvature was used to investigate non-equilibrium  steady-state quantum phase transitions~\cite{carollo2018uhlmann,leonforte2019uhlmann,carollo2020geometry}. 
Candeloro \textit{et al.} adopted this quantity as an asymptotic incompatibility
measure in multiparameter quantum estimation~\cite{candeloro}.
However, here we see that the quantum curvature is a natural generalization of the Berry curvature from pure to mixed states. Thus, we call it mixed-state Berry curvature.

Now we write the trade-off relation~\eqref{trade} in the following form as
\begin{align}\label{trade2}
	&{\cal F}_\beta\delta_\alpha^2+{\cal F}_\alpha\delta_\beta^2+2\delta_\alpha\delta_\beta\sqrt{{\cal F}_\alpha{\cal F}_\beta-|{\rm Im}\langle L_\alpha L_\beta\rangle|^2}\notag\\
	&\ge |{\rm Im}\langle L_\alpha L_\beta\rangle|^2,
\end{align}
where 
\begin{equation}
	\delta_\alpha^2={\cal F}_\alpha-F_\alpha.
\end{equation}
Then, from the definition of the quantum curvature~\eqref{bcmix}, we have 
\begin{align}\label{trade3}
	{\cal F}_\beta\delta_\alpha^2+{\cal F}_\alpha\delta_\beta^2+2\delta_\alpha\delta_\beta\sqrt{{\cal F}_\alpha{\cal F}_\beta-2\Omega_{\alpha\beta}^2}\ge 2\Omega_{\alpha\beta}^2.
\end{align}
Every entry in the above equation is clear. It includes two QFIs, two classical Fisher information and one quantum curvature. Similarly, the inequality~\eqref{trade222} can be written as
\begin{equation}\label{trade33}
	{{\cal F}_\alpha {\cal F}_\beta \geq 4 \Omega_{\alpha\beta}^2.}
\end{equation}
We see that the multiplication of two Fisher information is bounded below by four times the square of Berry curvature. For the case of zero quantum curvature, there is no trade-off relations between the two regrets in Eq.~\eqref{trade3} or two QFIs in Eq.~\eqref{trade33}. Thus, the quantum curvature plays a key role in the field of multi-parameter estimations.

{\em Quantum geometric tensor and spectral decomposition.---}The quantum curvature given above is closely related to the quantum geometric tensor(QGT)~\cite{qgt,preports,prbhou,zhouprb}. Let us first consider the QGT of a pure state $|\psi\rangle$ with parameters $\alpha$ and $\beta$. It is defined as
\begin{align}\label{qgtpure}
Q_{\alpha\beta}(\psi)&=\langle\partial_{\alpha}\psi|\partial_{\beta}\psi\rangle-\langle\partial_{\alpha}\psi|\psi\rangle\langle\psi|\partial_{\beta}\psi\rangle\notag\\
&=\langle\partial_{\alpha}\psi|\partial_{\beta}\psi\rangle+\langle\psi|\partial_{\alpha}\psi\rangle\langle\psi|\partial_{\beta}\psi\rangle.
\end{align}
Noting that
$\langle\partial_{\alpha}\psi|\psi\rangle$ is purely imaginary, we
have
\begin{equation}
\Omega_{\alpha\beta}(\psi)=-2\mathrm{Im}Q_{\alpha\beta}(\psi),
\end{equation}
where we have used Eq.~\eqref{bc}. For mixed states $\rho$, the QGT is given by
as
\begin{equation}
Q_{\alpha\beta}(\rho)=\frac{1}{4}\mathrm{Tr}\left(\rho L_{\alpha}L_{\beta}\right).
\end{equation}
Therefore, from the definition of quantum curvature for mixed states~\eqref{bcmix}, we obtain
\begin{equation}\label{relation}
\Omega_{\alpha\beta}(\rho)=-\frac{1}{2}\mathrm{Im}\,\mathrm{Tr}\left(\rho L_{\alpha}L_{\beta}\right)=-2\mathrm{Im}Q_{\alpha\beta}(\rho).
\end{equation}
The quantum curvature is the imaginary part of the QGT multiplied by minus two.

For a density matrix with full rank, the spectral decomposition of density matrix $\rho$ is written as
\begin{equation}
	\rho = \sum_{i=1}^{N} p_{i} |\psi_{i}\rangle\langle\psi_{i}|.
\end{equation}
where $N$ is the dimension of the Hilbert space. The QGT was given by~\cite{xgwang}
\begin{align}\label{qgtt}
	Q_{\alpha\beta}(\rho) 
	=&\frac{1}{4}\sum_{i=1}^{N}\frac{\left(\partial_{\alpha}p_{i}\right)\left(\partial_{\beta}p_{i}\right)}{p_{i}}\notag\\
	 +&\frac{1}{2}\sum_{i\neq j}\left[\frac{\left(p_{i}-p_{j}\right)^{2}}{\left(p_{i}+p_{j}\right)}{\cal R}_{ij}^{\alpha\beta}+{\rm i}\frac{\left(p_{i}-p_{j}\right)^{3}}{\left(p_{i}+p_{j}\right)^{2}}{\cal I}_{ij}^{\alpha\beta}\right],
\end{align}
where ${\cal R}_{ij}^{\alpha\beta}$ and ${\cal I}_{ij}^{\alpha\beta}$ are respectively the real and imaginary part of the double Wilczek-Zee connection 
\begin{align}\label{dwzc}
	{\cal A}_{ij}^{\alpha\beta}
	&=\langle\psi_{i}|\partial_{\alpha}\psi_{j}\rangle\langle\psi_{i}|\partial_{\beta}\psi_{j}\rangle^{*}\notag\\
	&=\langle\partial_\alpha\psi_{i}|\psi_{j}\rangle\langle\psi_{j}|\partial_\beta\psi_{i}\rangle.
\end{align}
Then from relation \eqref{relation} and the form of QGT \eqref{qgtt}, we finally obtain
\begin{align}
	\Omega_{\alpha\beta} (\rho)
	 &=-\sum_{i\neq j}\frac{\left(p_{i}-p_{j}\right)^{3}}{\left(p_{i}+p_{j}\right)^{2}}{\cal I}_{ij}^{\alpha\beta}\notag\\
	 &=-\sum_{i\neq j}\frac{\left(p_{i}-p_{j}\right)^{3}}{\left(p_{i}+p_{j}\right)^{2}}\text{Im}\left(\langle\partial_{\alpha}\psi_{i}|\psi_{j}\rangle\langle\psi_{j}|\partial_\beta\psi_{i}\rangle\right)
	 .
\end{align}
 Moreover, due to the fact ${\cal I}_{ij}^{\alpha\beta}=-{\cal I}_{ji}^{\alpha\beta}$, the above equation can be written as
\begin{align}\label{bcnew}
	\Omega_{\alpha\beta}(\rho) 
	=-2\sum_{i< j}\frac{\left(p_{i}-p_{j}\right)^{3}}{\left(p_{i}+p_{j}\right)^{2}}{\cal I}_{ij}^{\alpha\beta}.
\end{align}
We see that the quantum curvature is only related to the imaginary part of the double Wilczek-Zee connection.

We now study the quantum curvature of a non-full-rank density matrix, i.e., the rank $M$ of the density operator is less than the dimension $N$
of the Hilbert space.  The density matrix is written as
\begin{equation}
	\rho=\sum_{i=1}^{M}p_{i}|\psi_{i}\rangle\langle\psi_{i}|.
\end{equation}
With the above spectral decomposition, the QGT is given as 
\begin{align}\label{qgttttt}
	Q_{\alpha\beta}(\rho)  
	=&\frac{1}4\sum_{i=1}^M\frac{\left(\partial_{\alpha}p_{i}\right)\left(\partial_{\beta}p_{i}\right)}{p_{i}}+\sum_{i=1}^{M}p_{i}(\langle\partial_{\alpha}\psi_{i}|\partial_{\beta}\psi_{i}\rangle-{\cal R}_{ii}^{\alpha\beta})\notag\\
	& -2\sum_{i\neq j}^{M}\frac{p_{i}p_{j}}{p_{i}+p_{j}}\left[{\cal R}_{ij}^{\alpha\beta}+{\rm i}\frac{p_{i}-p_{j}}{p_{i}+p_{j}}{\cal I}_{ij}^{\alpha\beta}\right].
\end{align}
By using Eqs.~\eqref{qgtpure} and \eqref{dwzc}, the QGT can be written in the following form
\begin{align}
	Q_{\alpha\beta}(\rho)  
	=&\frac{1}4\sum_{i=1}^M\frac{\left(\partial_{\alpha}p_{i}\right)\left(\partial_{\beta}p_{i}\right)}{p_{i}}+\sum_{i=1}^{M}p_{i}Q_{\alpha\beta}(\psi_i)\notag\\
	& -2\sum_{i\neq j}^{M}\frac{p_{i}p_{j}}{p_{i}+p_{j}}\left[{\cal R}_{ij}^{\alpha\beta}+{\rm i}\frac{p_{i}-p_{j}}{p_{i}+p_{j}}{\cal I}_{ij}^{\alpha\beta}\right].
\end{align}

Therefore, for the density operators whose rank is less than the dimension
of the Hilbert space, we have
\begin{align}
\Omega_{\alpha\beta}(\rho) =\sum_{i=1}^{M}p_{i}\Omega_{\alpha\beta}(\psi_{i})+4\sum_{i\neq j}^{M}\frac{p_{i}p_{j}\left(p_{i}-p_{j}\right)}{\left(p_{i}+p_{j}\right)^{2}}{\cal I}^{\alpha\beta}_{ij}.
\end{align}
The first term in the above equation is just the average of Berry curvature of all pure states $|\psi_i\rangle$ in the spectral decomposition of the density matrix.
One can also simply define the quantum curvature for mixed states as
\begin{align}
	\Omega^\prime_{\alpha\beta}(\rho) =\sum_{i=1}^{M}p_{i}\Omega_{\alpha\beta}(\psi_{i}).
\end{align}
It is clear that both these quantum curvatures reduces to the Berry curvature of the pure state $|\psi_1\rangle$ when the rank of the density matrix becomes one.

{\em Example.---}As an example, we consider the pure state of a qubit with the following Bloch presentation
\begin{equation}
	\rho=\frac{1}{2}\left({I}+{\bm{n}}\cdot{\bm\sigma}\right)
\end{equation}
with ${\bm n}$ be a unit vector and ${\bm \sigma}$ be the Pauli matrix vector. Substituting the above equation into Eq.~\eqref{sld1} leads to
\begin{equation}
	L_{\alpha}=\left(\partial_{\alpha}{\bm n}\right)\cdot{\bm \sigma}.
\end{equation}
For two SLDs $L_{\alpha}$ and $L_\beta$, the expectation of the commutator
$[L_{\alpha},L_{\beta}]$ on state $\rho$ can be obtained as
\begin{equation}
	\langle[L_{\alpha},L_{\beta}]\rangle=\mathrm{Tr}\left[\rho[L_{\alpha},L_{\beta}]\right]=2i{\bm n}\cdot\partial_{\alpha}{\bm n}\times\partial_{\beta}{\bm n},
\end{equation}
where we have used the identity
\begin{equation}
	{\rm Tr}(({\bm C}\cdot {\bm \sigma})[({\bm A}\cdot {\bm \sigma}),({\bm B}\cdot {\bm \sigma})])=4i{\bm C}\cdot ({\bm A}\times {\bm B}).
\end{equation}

Then, the Berry curvature is given in the following form
\begin{equation}
	\Omega_{\alpha\beta}=\frac{i}{4}\langle[L_{\alpha},L_{\beta}]\rangle=-\frac{1}2{\bm n}\cdot\partial_{\alpha}{\bm n}\times\partial_{\beta}{\bm n}.
\end{equation}
The geometric meaning is clear. The quantum curvature is 
the volume of the parallelepiped spanned by the three vectors ${\bm n}$, $\partial_{\alpha}{\bm n}$, and $\partial_{\beta}{\bm n}$ up to a constant.

We know investigate a mixed state of a qubit with the Bloch representation
\begin{equation}
	\rho=\frac{1}{2}\left(I+{\bm r}\cdot{\bm \sigma}\right),
\end{equation}
where ${\bm r}$ is a 3-dimensional real vector whose length $r$
is less than $1$. From Eq.~\eqref{bcnew}, the quantum curvature is given by
\begin{align}
	\Omega_{\alpha\beta}(\rho) 
	=-2\left(p_{1}-p_{2}\right)^{3}\text{Im}\left(\langle\partial_{\alpha}\psi_{1}|\psi_{2}\rangle\langle\partial_{\beta}\psi_{1}|\psi_{2}\rangle^{*}\right),
\end{align}
where we have used the fact $p_1+p_2=1$. Then one step further leads to
\begin{align}
	\Omega_{\alpha\beta}(\rho) 
	& =-2\left(p_{1}-p_{2}\right)^{3}\text{Im}\left(\langle\partial_{\alpha}\psi_{1}|\psi_{2}\rangle\langle\psi_{2}|\partial_{\beta}\psi_{1}\rangle\right)\\
	& =-2\left(p_{1}-p_{2}\right)^{3}\notag\\
	\times& \text{Im}\left(\langle\partial_{\alpha}\psi_{1}|\partial_{\beta}\psi_{1}\rangle-\langle\partial_{\alpha}\psi_{1}|\psi_{1}\rangle\langle\psi_{1}|\partial_{\beta}\psi_{1}\rangle\right),
\end{align}
where the completeness relation was used.
As $\langle\partial_{\alpha}\psi_{1}|\psi_{1}\rangle$ and $\langle\partial_{\beta}\psi_{1}|\psi_{1}\rangle$
are all purely imaginary, we finally have a simple form
\begin{equation}\label{bcqubit}
\Omega_{\alpha\beta}(\rho)=-2\left(p_{1}-p_{2}\right)^{3}\text{Im}\left(\langle\partial_{\alpha}\psi_{1}|\partial_{\beta}\psi_{1}\rangle\right).
\end{equation}
With the Bloch representation of density operators of a qubit, we
have $p_{1}=(1-r)/2$ and $p_{2}=(1+r)/2$ and thus
\begin{equation}
\left(p_{1}-p_{2}\right)^{3}=-r^{3}.
\end{equation}
Then, from definition \eqref{bc} and Eq. \eqref{bcmix}, we get  
\begin{equation}
\Omega_{\alpha\beta}(\rho)=-r^{3}\Omega_{\alpha\beta}(\psi_{1}).
\end{equation}
The  quantum curvature is proportional to the Berry curvature for pure state $|\psi_{1}\rangle$. When $r=0$, the quantum curvature vanishes as the corresponding state is the completely mixed state.

{\em Conclusions.---}In conclusion, we have proposed a mixed-state quantum curvature based on the SLDs in the field of quantum precision measurement. For full-rank and non-full-rank density matrices , we provided explicit expressions of quantum curvature after the spectral decomposition. Our work extended the quantum curvature to the case of mixed states and the quantum curvature was found to be a key role  in the field of multi-parameter estimations. And this concept is expected to play important roles in the study of fundamental problems of quantum physics.

\section*{Acknowledgements}
This work is supported by the Innovation Pro-gram
for Quantum Science and Technology (Grant No.2024ZD0301000), Science Challenge Project (Grant No.TZ2025017), and the Science Foundation of Zhejiang Sci-Tech University (Grants No.23062088-Y).

\bibliography{myref}

\begin{thebibliography}{28}%
\makeatletter
\providecommand \@ifxundefined [1]{%
 \@ifx{#1\undefined}
}%
\providecommand \@ifnum [1]{%
 \ifnum #1\expandafter \@firstoftwo
 \else \expandafter \@secondoftwo
 \fi
}%
\providecommand \@ifx [1]{%
 \ifx #1\expandafter \@firstoftwo
 \else \expandafter \@secondoftwo
 \fi
}%
\providecommand \natexlab [1]{#1}%
\providecommand \enquote  [1]{``#1''}%
\providecommand \bibnamefont  [1]{#1}%
\providecommand \bibfnamefont [1]{#1}%
\providecommand \citenamefont [1]{#1}%
\providecommand \href@noop [0]{\@secondoftwo}%
\providecommand \href [0]{\begingroup \@sanitize@url \@href}%
\providecommand \@href[1]{\@@startlink{#1}\@@href}%
\providecommand \@@href[1]{\endgroup#1\@@endlink}%
\providecommand \@sanitize@url [0]{\catcode `\\12\catcode `\$12\catcode
  `\&12\catcode `\#12\catcode `\^12\catcode `\_12\catcode `\%12\relax}%
\providecommand \@@startlink[1]{}%
\providecommand \@@endlink[0]{}%
\providecommand \url  [0]{\begingroup\@sanitize@url \@url }%
\providecommand \@url [1]{\endgroup\@href {#1}{\urlprefix }}%
\providecommand \urlprefix  [0]{URL }%
\providecommand \Eprint [0]{\href }%
\providecommand \doibase [0]{https://doi.org/}%
\providecommand \selectlanguage [0]{\@gobble}%
\providecommand \bibinfo  [0]{\@secondoftwo}%
\providecommand \bibfield  [0]{\@secondoftwo}%
\providecommand \translation [1]{[#1]}%
\providecommand \BibitemOpen [0]{}%
\providecommand \bibitemStop [0]{}%
\providecommand \bibitemNoStop [0]{.\EOS\space}%
\providecommand \EOS [0]{\spacefactor3000\relax}%
\providecommand \BibitemShut  [1]{\csname bibitem#1\endcsname}%
\let\auto@bib@innerbib\@empty
\bibitem [{\citenamefont {Berry}(1984)}]{Berry1984}%
  \BibitemOpen
  \bibfield  {author} {\bibinfo {author} {\bibfnamefont {M.~V.}\ \bibnamefont
  {Berry}},\ }\bibfield  {title} {\bibinfo {title} {Quantal phase factors
  accompanying adiabatic changes},\ }\href@noop {} {\bibfield  {journal}
  {\bibinfo  {journal} {Proceedings of the Royal Society of London. A.
  Mathematical and Physical Sciences}\ }\textbf {\bibinfo {volume} {392}},\
  \bibinfo {pages} {45} (\bibinfo {year} {1984})}\BibitemShut {NoStop}%
\bibitem [{\citenamefont {Son}\ and\ \citenamefont {Yamamoto}(2012)}]{Son2012}%
  \BibitemOpen
  \bibfield  {author} {\bibinfo {author} {\bibfnamefont {D.~T.}\ \bibnamefont
  {Son}}\ and\ \bibinfo {author} {\bibfnamefont {N.}~\bibnamefont {Yamamoto}},\
  }\bibfield  {title} {\bibinfo {title} {Berry curvature, triangle anomalies,
  and the chiral magnetic effect in fermi liquids},\ }\href@noop {} {\bibfield
  {journal} {\bibinfo  {journal} {Phys. Rev. Lett.}\ }\textbf {\bibinfo
  {volume} {109}},\ \bibinfo {pages} {181602} (\bibinfo {year}
  {2012})}\BibitemShut {NoStop}%
\bibitem [{\citenamefont {Murakami}\ \emph {et~al.}(2003)\citenamefont
  {Murakami}, \citenamefont {Nagaosa},\ and\ \citenamefont {Zhang}}]{SCZhang}%
  \BibitemOpen
  \bibfield  {author} {\bibinfo {author} {\bibfnamefont {S.}~\bibnamefont
  {Murakami}}, \bibinfo {author} {\bibfnamefont {N.}~\bibnamefont {Nagaosa}},\
  and\ \bibinfo {author} {\bibfnamefont {S.-C.}\ \bibnamefont {Zhang}},\
  }\bibfield  {title} {\bibinfo {title} {Dissipationless quantum spin current
  at room temperature},\ }\href@noop {} {\bibfield  {journal} {\bibinfo
  {journal} {Science}\ }\textbf {\bibinfo {volume} {301}},\ \bibinfo {pages}
  {1348} (\bibinfo {year} {2003})}\BibitemShut {NoStop}%
\bibitem [{\citenamefont {Nagaosa}\ \emph {et~al.}(2010)\citenamefont
  {Nagaosa}, \citenamefont {Sinova}, \citenamefont {Onoda}, \citenamefont
  {MacDonald},\ and\ \citenamefont {Ong}}]{AHall}%
  \BibitemOpen
  \bibfield  {author} {\bibinfo {author} {\bibfnamefont {N.}~\bibnamefont
  {Nagaosa}}, \bibinfo {author} {\bibfnamefont {J.}~\bibnamefont {Sinova}},
  \bibinfo {author} {\bibfnamefont {S.}~\bibnamefont {Onoda}}, \bibinfo
  {author} {\bibfnamefont {A.~H.}\ \bibnamefont {MacDonald}},\ and\ \bibinfo
  {author} {\bibfnamefont {N.~P.}\ \bibnamefont {Ong}},\ }\bibfield  {title}
  {\bibinfo {title} {Anomalous hall effect},\ }\href
  {https://doi.org/10.1103/RevModPhys.82.1539} {\bibfield  {journal} {\bibinfo
  {journal} {Rev. Mod. Phys.}\ }\textbf {\bibinfo {volume} {82}},\ \bibinfo
  {pages} {1539} (\bibinfo {year} {2010})}\BibitemShut {NoStop}%
\bibitem [{\citenamefont {Dau}\ \emph {et~al.}(2019)\citenamefont {Dau},
  \citenamefont {Vergnaud}, \citenamefont {Marty}, \citenamefont {Beign{\'e}},
  \citenamefont {Gambarelli}, \citenamefont {Maurel}, \citenamefont {Journot},
  \citenamefont {Hyot}, \citenamefont {Guillet}, \citenamefont {Gr{\'e}vin}
  \emph {et~al.}}]{Nernst}%
  \BibitemOpen
  \bibfield  {author} {\bibinfo {author} {\bibfnamefont {M.~T.}\ \bibnamefont
  {Dau}}, \bibinfo {author} {\bibfnamefont {C.}~\bibnamefont {Vergnaud}},
  \bibinfo {author} {\bibfnamefont {A.}~\bibnamefont {Marty}}, \bibinfo
  {author} {\bibfnamefont {C.}~\bibnamefont {Beign{\'e}}}, \bibinfo {author}
  {\bibfnamefont {S.}~\bibnamefont {Gambarelli}}, \bibinfo {author}
  {\bibfnamefont {V.}~\bibnamefont {Maurel}}, \bibinfo {author} {\bibfnamefont
  {T.}~\bibnamefont {Journot}}, \bibinfo {author} {\bibfnamefont
  {B.}~\bibnamefont {Hyot}}, \bibinfo {author} {\bibfnamefont {T.}~\bibnamefont
  {Guillet}}, \bibinfo {author} {\bibfnamefont {B.}~\bibnamefont {Gr{\'e}vin}},
  \emph {et~al.},\ }\bibfield  {title} {\bibinfo {title} {The valley {Nernst}
  effect in {WSe2}},\ }\href@noop {} {\bibfield  {journal} {\bibinfo  {journal}
  {Nat. Commun.}\ }\textbf {\bibinfo {volume} {10}},\ \bibinfo {pages} {5796}
  (\bibinfo {year} {2019})}\BibitemShut {NoStop}%
\bibitem [{\citenamefont {Sogabe}\ and\ \citenamefont {Yin}(2025)}]{bcsc}%
  \BibitemOpen
  \bibfield  {author} {\bibinfo {author} {\bibfnamefont {N.}~\bibnamefont
  {Sogabe}}\ and\ \bibinfo {author} {\bibfnamefont {Y.}~\bibnamefont {Yin}},\
  }\bibfield  {title} {\bibinfo {title} {Berry curvature and spin-one color
  superconductivity},\ }\href@noop {} {\bibfield  {journal} {\bibinfo
  {journal} {Phys. Rev. Lett.}\ }\textbf {\bibinfo {volume} {134}},\ \bibinfo
  {pages} {171903} (\bibinfo {year} {2025})}\BibitemShut {NoStop}%
\bibitem [{\citenamefont {Chen}\ \emph {et~al.}(2023)\citenamefont {Chen},
  \citenamefont {Liu}, \citenamefont {Wu}, \citenamefont {Su}, \citenamefont
  {Ding}, \citenamefont {Qin}, \citenamefont {Wang}, \citenamefont {Zhang},
  \citenamefont {He}, \citenamefont {Wang}, \citenamefont {Lu}, \citenamefont
  {Li}, \citenamefont {Sanders}, \citenamefont {Liu},\ and\ \citenamefont
  {Pan}}]{SandersPan}%
  \BibitemOpen
  \bibfield  {author} {\bibinfo {author} {\bibfnamefont {C.}~\bibnamefont
  {Chen}}, \bibinfo {author} {\bibfnamefont {R.-Z.}\ \bibnamefont {Liu}},
  \bibinfo {author} {\bibfnamefont {J.}~\bibnamefont {Wu}}, \bibinfo {author}
  {\bibfnamefont {Z.-E.}\ \bibnamefont {Su}}, \bibinfo {author} {\bibfnamefont
  {X.}~\bibnamefont {Ding}}, \bibinfo {author} {\bibfnamefont {J.}~\bibnamefont
  {Qin}}, \bibinfo {author} {\bibfnamefont {L.}~\bibnamefont {Wang}}, \bibinfo
  {author} {\bibfnamefont {W.-W.}\ \bibnamefont {Zhang}}, \bibinfo {author}
  {\bibfnamefont {Y.}~\bibnamefont {He}}, \bibinfo {author} {\bibfnamefont
  {X.-L.}\ \bibnamefont {Wang}}, \bibinfo {author} {\bibfnamefont {C.-Y.}\
  \bibnamefont {Lu}}, \bibinfo {author} {\bibfnamefont {L.}~\bibnamefont {Li}},
  \bibinfo {author} {\bibfnamefont {B.~C.}\ \bibnamefont {Sanders}}, \bibinfo
  {author} {\bibfnamefont {X.-J.}\ \bibnamefont {Liu}},\ and\ \bibinfo {author}
  {\bibfnamefont {J.-W.}\ \bibnamefont {Pan}},\ }\bibfield  {title} {\bibinfo
  {title} {Berry curvature and bulk-boundary correspondence from transport
  measurement for photonic chern bands},\ }\href
  {https://doi.org/10.1103/PhysRevLett.131.133601} {\bibfield  {journal}
  {\bibinfo  {journal} {Phys. Rev. Lett.}\ }\textbf {\bibinfo {volume} {131}},\
  \bibinfo {pages} {133601} (\bibinfo {year} {2023})}\BibitemShut {NoStop}%
\bibitem [{\citenamefont {Bac}\ \emph {et~al.}(2025)\citenamefont {Bac},
  \citenamefont {Le~Mardel\'e}, \citenamefont {Wang}, \citenamefont {Ozerov},
  \citenamefont {Yoshimura}, \citenamefont {Mohelsk\'y}, \citenamefont {Sun},
  \citenamefont {Piot}, \citenamefont {Wimmer}, \citenamefont {Ney},
  \citenamefont {Orlova}, \citenamefont {Zhukovskyi}, \citenamefont {Bauer},
  \citenamefont {Springholz}, \citenamefont {Liu}, \citenamefont {Orlita},
  \citenamefont {Park}, \citenamefont {Hsu},\ and\ \citenamefont
  {Assaf}}]{bcexp1}%
  \BibitemOpen
  \bibfield  {author} {\bibinfo {author} {\bibfnamefont {S.-K.}\ \bibnamefont
  {Bac}}, \bibinfo {author} {\bibfnamefont {F.}~\bibnamefont {Le~Mardel\'e}},
  \bibinfo {author} {\bibfnamefont {J.}~\bibnamefont {Wang}}, \bibinfo {author}
  {\bibfnamefont {M.}~\bibnamefont {Ozerov}}, \bibinfo {author} {\bibfnamefont
  {K.}~\bibnamefont {Yoshimura}}, \bibinfo {author} {\bibfnamefont
  {I.}~\bibnamefont {Mohelsk\'y}}, \bibinfo {author} {\bibfnamefont
  {X.}~\bibnamefont {Sun}}, \bibinfo {author} {\bibfnamefont {B.~A.}\
  \bibnamefont {Piot}}, \bibinfo {author} {\bibfnamefont {S.}~\bibnamefont
  {Wimmer}}, \bibinfo {author} {\bibfnamefont {A.}~\bibnamefont {Ney}},
  \bibinfo {author} {\bibfnamefont {T.}~\bibnamefont {Orlova}}, \bibinfo
  {author} {\bibfnamefont {M.}~\bibnamefont {Zhukovskyi}}, \bibinfo {author}
  {\bibfnamefont {G.}~\bibnamefont {Bauer}}, \bibinfo {author} {\bibfnamefont
  {G.}~\bibnamefont {Springholz}}, \bibinfo {author} {\bibfnamefont
  {X.}~\bibnamefont {Liu}}, \bibinfo {author} {\bibfnamefont {M.}~\bibnamefont
  {Orlita}}, \bibinfo {author} {\bibfnamefont {K.}~\bibnamefont {Park}},
  \bibinfo {author} {\bibfnamefont {Y.-T.}\ \bibnamefont {Hsu}},\ and\ \bibinfo
  {author} {\bibfnamefont {B.~A.}\ \bibnamefont {Assaf}},\ }\bibfield  {title}
  {\bibinfo {title} {Probing berry curvature in magnetic topological insulators
  through resonant infrared magnetic circular dichroism},\ }\href@noop {}
  {\bibfield  {journal} {\bibinfo  {journal} {Phys. Rev. Lett.}\ }\textbf
  {\bibinfo {volume} {134}},\ \bibinfo {pages} {016601} (\bibinfo {year}
  {2025})}\BibitemShut {NoStop}%
\bibitem [{\citenamefont {Gritsev}\ and\ \citenamefont
  {Polkovnikov}(2012)}]{gritsev2012dynamical}%
  \BibitemOpen
  \bibfield  {author} {\bibinfo {author} {\bibfnamefont {V.}~\bibnamefont
  {Gritsev}}\ and\ \bibinfo {author} {\bibfnamefont {A.}~\bibnamefont
  {Polkovnikov}},\ }\bibfield  {title} {\bibinfo {title} {Dynamical quantum
  hall effect in the parameter space},\ }\href@noop {} {\bibfield  {journal}
  {\bibinfo  {journal} {Proceedings of the National Academy of Sciences}\
  }\textbf {\bibinfo {volume} {109}},\ \bibinfo {pages} {6457} (\bibinfo {year}
  {2012})}\BibitemShut {NoStop}%
\bibitem [{\citenamefont {Sj{\"o}qvist}\ \emph {et~al.}(2000)\citenamefont
  {Sj{\"o}qvist}, \citenamefont {Pati}, \citenamefont {Ekert}, \citenamefont
  {Anandan}, \citenamefont {Ericsson}, \citenamefont {Oi},\ and\ \citenamefont
  {Vedral}}]{sjoqvist2000geometric}%
  \BibitemOpen
  \bibfield  {author} {\bibinfo {author} {\bibfnamefont {E.}~\bibnamefont
  {Sj{\"o}qvist}}, \bibinfo {author} {\bibfnamefont {A.~K.}\ \bibnamefont
  {Pati}}, \bibinfo {author} {\bibfnamefont {A.}~\bibnamefont {Ekert}},
  \bibinfo {author} {\bibfnamefont {J.~S.}\ \bibnamefont {Anandan}}, \bibinfo
  {author} {\bibfnamefont {M.}~\bibnamefont {Ericsson}}, \bibinfo {author}
  {\bibfnamefont {D.~K.}\ \bibnamefont {Oi}},\ and\ \bibinfo {author}
  {\bibfnamefont {V.}~\bibnamefont {Vedral}},\ }\bibfield  {title} {\bibinfo
  {title} {Geometric phases for mixed states in interferometry},\ }\href@noop
  {} {\bibfield  {journal} {\bibinfo  {journal} {Physical Review Letters}\
  }\textbf {\bibinfo {volume} {85}},\ \bibinfo {pages} {2845} (\bibinfo {year}
  {2000})}\BibitemShut {NoStop}%
\bibitem [{\citenamefont {Lu}\ and\ \citenamefont {Wang}(2021)}]{luwang}%
  \BibitemOpen
  \bibfield  {author} {\bibinfo {author} {\bibfnamefont {X.-M.}\ \bibnamefont
  {Lu}}\ and\ \bibinfo {author} {\bibfnamefont {X.}~\bibnamefont {Wang}},\
  }\bibfield  {title} {\bibinfo {title} {Incorporating {Heisenberg's}
  uncertainty principle into quantum multiparameter estimation},\ }\href
  {https://doi.org/10.1103/PhysRevLett.126.120503} {\bibfield  {journal}
  {\bibinfo  {journal} {Phys. Rev. Lett.}\ }\textbf {\bibinfo {volume} {126}},\
  \bibinfo {pages} {120503} (\bibinfo {year} {2021})}\BibitemShut {NoStop}%
\bibitem [{\citenamefont {Helstrom}(1969)}]{helstrom1969}%
  \BibitemOpen
  \bibfield  {author} {\bibinfo {author} {\bibfnamefont {C.~W.}\ \bibnamefont
  {Helstrom}},\ }\bibfield  {title} {\bibinfo {title} {Quantum detection and
  estimation theory},\ }\href
  {https://api.semanticscholar.org/CorpusID:12758217} {\bibfield  {journal}
  {\bibinfo  {journal} {Journal of Statistical Physics}\ }\textbf {\bibinfo
  {volume} {1}},\ \bibinfo {pages} {231} (\bibinfo {year} {1969})}\BibitemShut
  {NoStop}%
\bibitem [{\citenamefont {Holevo}(2011)}]{holevo2011probabilistic}%
  \BibitemOpen
  \bibfield  {author} {\bibinfo {author} {\bibfnamefont {A.~S.}\ \bibnamefont
  {Holevo}},\ }\href@noop {} {\emph {\bibinfo {title} {Probabilistic and
  statistical aspects of quantum theory}}},\ Vol.~\bibinfo {volume} {1}\
  (\bibinfo  {publisher} {Springer Science \& Business Media},\ \bibinfo {year}
  {2011})\BibitemShut {NoStop}%
\bibitem [{\citenamefont {Zhang}\ \emph {et~al.}(2023)\citenamefont {Zhang},
  \citenamefont {Lu}, \citenamefont {Liu}, \citenamefont {Ding},\ and\
  \citenamefont {Wang}}]{xingyu}%
  \BibitemOpen
  \bibfield  {author} {\bibinfo {author} {\bibfnamefont {X.}~\bibnamefont
  {Zhang}}, \bibinfo {author} {\bibfnamefont {X.-M.}\ \bibnamefont {Lu}},
  \bibinfo {author} {\bibfnamefont {J.}~\bibnamefont {Liu}}, \bibinfo {author}
  {\bibfnamefont {W.}~\bibnamefont {Ding}},\ and\ \bibinfo {author}
  {\bibfnamefont {X.}~\bibnamefont {Wang}},\ }\bibfield  {title} {\bibinfo
  {title} {Direct measurement of quantum fisher information},\ }\href
  {https://doi.org/10.1103/PhysRevA.107.012414} {\bibfield  {journal} {\bibinfo
   {journal} {Phys. Rev. A}\ }\textbf {\bibinfo {volume} {107}},\ \bibinfo
  {pages} {012414} (\bibinfo {year} {2023})}\BibitemShut {NoStop}%
\bibitem [{\citenamefont {Ding}\ \emph {et~al.}(2024)\citenamefont {Ding},
  \citenamefont {Zhang}, \citenamefont {Liu}, \citenamefont {Chen},\ and\
  \citenamefont {Wang}}]{wenkuiprb}%
  \BibitemOpen
  \bibfield  {author} {\bibinfo {author} {\bibfnamefont {W.}~\bibnamefont
  {Ding}}, \bibinfo {author} {\bibfnamefont {X.}~\bibnamefont {Zhang}},
  \bibinfo {author} {\bibfnamefont {J.}~\bibnamefont {Liu}}, \bibinfo {author}
  {\bibfnamefont {A.}~\bibnamefont {Chen}},\ and\ \bibinfo {author}
  {\bibfnamefont {X.}~\bibnamefont {Wang}},\ }\bibfield  {title} {\bibinfo
  {title} {Quantum dynamic response-based {NV-diamond} magnetometry: Robustness
  to decoherence and applications in motion detection of magnetic
  nanoparticles},\ }\href {https://doi.org/10.1103/PhysRevB.110.045202}
  {\bibfield  {journal} {\bibinfo  {journal} {Phys. Rev. B}\ }\textbf {\bibinfo
  {volume} {110}},\ \bibinfo {pages} {045202} (\bibinfo {year}
  {2024})}\BibitemShut {NoStop}%
\bibitem [{\citenamefont {Conlon}\ \emph {et~al.}(2023)\citenamefont {Conlon},
  \citenamefont {Vogl}, \citenamefont {Marciniak}, \citenamefont {Pogorelov},
  \citenamefont {Yung}, \citenamefont {Eilenberger}, \citenamefont {Berry},
  \citenamefont {Santana}, \citenamefont {Blatt}, \citenamefont {Monz} \emph
  {et~al.}}]{conlon2023approaching}%
  \BibitemOpen
  \bibfield  {author} {\bibinfo {author} {\bibfnamefont {L.~O.}\ \bibnamefont
  {Conlon}}, \bibinfo {author} {\bibfnamefont {T.}~\bibnamefont {Vogl}},
  \bibinfo {author} {\bibfnamefont {C.~D.}\ \bibnamefont {Marciniak}}, \bibinfo
  {author} {\bibfnamefont {I.}~\bibnamefont {Pogorelov}}, \bibinfo {author}
  {\bibfnamefont {S.~K.}\ \bibnamefont {Yung}}, \bibinfo {author}
  {\bibfnamefont {F.}~\bibnamefont {Eilenberger}}, \bibinfo {author}
  {\bibfnamefont {D.~W.}\ \bibnamefont {Berry}}, \bibinfo {author}
  {\bibfnamefont {F.~S.}\ \bibnamefont {Santana}}, \bibinfo {author}
  {\bibfnamefont {R.}~\bibnamefont {Blatt}}, \bibinfo {author} {\bibfnamefont
  {T.}~\bibnamefont {Monz}}, \emph {et~al.},\ }\bibfield  {title} {\bibinfo
  {title} {Approaching optimal entangling collective measurements on quantum
  computing platforms},\ }\href@noop {} {\bibfield  {journal} {\bibinfo
  {journal} {Nature Physics}\ }\textbf {\bibinfo {volume} {19}},\ \bibinfo
  {pages} {351} (\bibinfo {year} {2023})}\BibitemShut {NoStop}%
\bibitem [{\citenamefont {Xia}\ \emph {et~al.}(2023)\citenamefont {Xia},
  \citenamefont {Huang}, \citenamefont {Li}, \citenamefont {Wang},\ and\
  \citenamefont {Zeng}}]{xia2023toward}%
  \BibitemOpen
  \bibfield  {author} {\bibinfo {author} {\bibfnamefont {B.}~\bibnamefont
  {Xia}}, \bibinfo {author} {\bibfnamefont {J.}~\bibnamefont {Huang}}, \bibinfo
  {author} {\bibfnamefont {H.}~\bibnamefont {Li}}, \bibinfo {author}
  {\bibfnamefont {H.}~\bibnamefont {Wang}},\ and\ \bibinfo {author}
  {\bibfnamefont {G.}~\bibnamefont {Zeng}},\ }\bibfield  {title} {\bibinfo
  {title} {Toward incompatible quantum limits on multiparameter estimation},\
  }\href@noop {} {\bibfield  {journal} {\bibinfo  {journal} {Nature
  Communications}\ }\textbf {\bibinfo {volume} {14}},\ \bibinfo {pages} {1021}
  (\bibinfo {year} {2023})}\BibitemShut {NoStop}%
\bibitem [{\citenamefont {Uhlmann}(1986)}]{uhlmann1986parallel}%
  \BibitemOpen
  \bibfield  {author} {\bibinfo {author} {\bibfnamefont {A.}~\bibnamefont
  {Uhlmann}},\ }\bibfield  {title} {\bibinfo {title} {Parallel transport and
  “quantum holonomy” along density operators},\ }\href@noop {} {\bibfield
  {journal} {\bibinfo  {journal} {Reports on Mathematical Physics}\ }\textbf
  {\bibinfo {volume} {24}},\ \bibinfo {pages} {229} (\bibinfo {year}
  {1986})}\BibitemShut {NoStop}%
\bibitem [{\citenamefont {Carollo}\ \emph {et~al.}(2019)\citenamefont
  {Carollo}, \citenamefont {Spagnolo}, \citenamefont {Dubkov},\ and\
  \citenamefont {Valenti}}]{carollo2019quantumness}%
  \BibitemOpen
  \bibfield  {author} {\bibinfo {author} {\bibfnamefont {A.}~\bibnamefont
  {Carollo}}, \bibinfo {author} {\bibfnamefont {B.}~\bibnamefont {Spagnolo}},
  \bibinfo {author} {\bibfnamefont {A.~A.}\ \bibnamefont {Dubkov}},\ and\
  \bibinfo {author} {\bibfnamefont {D.}~\bibnamefont {Valenti}},\ }\bibfield
  {title} {\bibinfo {title} {On quantumness in multi-parameter quantum
  estimation},\ }\href@noop {} {\bibfield  {journal} {\bibinfo  {journal}
  {Journal of Statistical Mechanics: Theory and Experiment}\ }\textbf {\bibinfo
  {volume} {2019}},\ \bibinfo {pages} {094010} (\bibinfo {year}
  {2019})}\BibitemShut {NoStop}%
\bibitem [{\citenamefont {Carollo}\ \emph {et~al.}(2018)\citenamefont
  {Carollo}, \citenamefont {Spagnolo},\ and\ \citenamefont
  {Valenti}}]{carollo2018uhlmann}%
  \BibitemOpen
  \bibfield  {author} {\bibinfo {author} {\bibfnamefont {A.}~\bibnamefont
  {Carollo}}, \bibinfo {author} {\bibfnamefont {B.}~\bibnamefont {Spagnolo}},\
  and\ \bibinfo {author} {\bibfnamefont {D.}~\bibnamefont {Valenti}},\
  }\bibfield  {title} {\bibinfo {title} {Uhlmann curvature in dissipative phase
  transitions},\ }\href@noop {} {\bibfield  {journal} {\bibinfo  {journal}
  {Scientific reports}\ }\textbf {\bibinfo {volume} {8}},\ \bibinfo {pages}
  {9852} (\bibinfo {year} {2018})}\BibitemShut {NoStop}%
\bibitem [{\citenamefont {Leonforte}\ \emph {et~al.}(2019)\citenamefont
  {Leonforte}, \citenamefont {Valenti}, \citenamefont {Spagnolo},\ and\
  \citenamefont {Carollo}}]{leonforte2019uhlmann}%
  \BibitemOpen
  \bibfield  {author} {\bibinfo {author} {\bibfnamefont {L.}~\bibnamefont
  {Leonforte}}, \bibinfo {author} {\bibfnamefont {D.}~\bibnamefont {Valenti}},
  \bibinfo {author} {\bibfnamefont {B.}~\bibnamefont {Spagnolo}},\ and\
  \bibinfo {author} {\bibfnamefont {A.}~\bibnamefont {Carollo}},\ }\bibfield
  {title} {\bibinfo {title} {Uhlmann number in translational invariant
  systems},\ }\href@noop {} {\bibfield  {journal} {\bibinfo  {journal}
  {Scientific Reports}\ }\textbf {\bibinfo {volume} {9}},\ \bibinfo {pages}
  {9106} (\bibinfo {year} {2019})}\BibitemShut {NoStop}%
\bibitem [{\citenamefont {Carollo}\ \emph {et~al.}(2020)\citenamefont
  {Carollo}, \citenamefont {Valenti},\ and\ \citenamefont
  {Spagnolo}}]{carollo2020geometry}%
  \BibitemOpen
  \bibfield  {author} {\bibinfo {author} {\bibfnamefont {A.}~\bibnamefont
  {Carollo}}, \bibinfo {author} {\bibfnamefont {D.}~\bibnamefont {Valenti}},\
  and\ \bibinfo {author} {\bibfnamefont {B.}~\bibnamefont {Spagnolo}},\
  }\bibfield  {title} {\bibinfo {title} {Geometry of quantum phase
  transitions},\ }\href@noop {} {\bibfield  {journal} {\bibinfo  {journal}
  {Physics Reports}\ }\textbf {\bibinfo {volume} {838}},\ \bibinfo {pages} {1}
  (\bibinfo {year} {2020})}\BibitemShut {NoStop}%
\bibitem [{\citenamefont {Candeloro}\ \emph {et~al.}()\citenamefont
  {Candeloro}, \citenamefont {Paris},\ and\ \citenamefont
  {Genoni}}]{candeloro}%
  \BibitemOpen
  \bibfield  {author} {\bibinfo {author} {\bibfnamefont {A.}~\bibnamefont
  {Candeloro}}, \bibinfo {author} {\bibfnamefont {M.~G.~A.}\ \bibnamefont
  {Paris}},\ and\ \bibinfo {author} {\bibfnamefont {M.~G.}\ \bibnamefont
  {Genoni}},\ }\bibfield  {title} {\bibinfo {title} {On the properties of the
  asymptotic incompatibility measure in multiparameter quantum estimation},\
  }\href@noop {} {\bibinfo  {journal} {arXiv:2107.13426.v2}\ }\BibitemShut
  {NoStop}%
\bibitem [{\citenamefont {Cheng}(2010)}]{qgt}%
  \BibitemOpen
\bibfield  {journal} {  }\bibfield  {author} {\bibinfo {author} {\bibfnamefont
  {R.}~\bibnamefont {Cheng}},\ }\bibfield  {title} {\bibinfo {title} {Quantum
  geometric tensor ({Fubini-Study} metric) in simple quantum system: A
  pedagogical introduction},\ }\href@noop {} {\bibfield  {journal} {\bibinfo
  {journal} {arXiv:1012.1337}\ } (\bibinfo {year} {2010})}\BibitemShut
  {NoStop}%
\bibitem [{\citenamefont {Kolodrubetz}\ \emph {et~al.}(2017)\citenamefont
  {Kolodrubetz}, \citenamefont {Sels}, \citenamefont {Mehta},\ and\
  \citenamefont {Polkovnikov}}]{preports}%
  \BibitemOpen
  \bibfield  {author} {\bibinfo {author} {\bibfnamefont {M.}~\bibnamefont
  {Kolodrubetz}}, \bibinfo {author} {\bibfnamefont {D.}~\bibnamefont {Sels}},
  \bibinfo {author} {\bibfnamefont {P.}~\bibnamefont {Mehta}},\ and\ \bibinfo
  {author} {\bibfnamefont {A.}~\bibnamefont {Polkovnikov}},\ }\bibfield
  {title} {\bibinfo {title} {Geometry and non-adiabatic response in quantum and
  classical systems},\ }\href {https://doi.org/10.1016/j.physrep.2017.07.001}
  {\bibfield  {journal} {\bibinfo  {journal} {Physics Reports}\ }\textbf
  {\bibinfo {volume} {697}},\ \bibinfo {pages} {1} (\bibinfo {year}
  {2017})}\BibitemShut {NoStop}%
\bibitem [{\citenamefont {Hou}\ \emph {et~al.}(2024)\citenamefont {Hou},
  \citenamefont {Zhou}, \citenamefont {Wang}, \citenamefont {Guo},\ and\
  \citenamefont {Chien}}]{prbhou}%
  \BibitemOpen
  \bibfield  {author} {\bibinfo {author} {\bibfnamefont {X.-Y.}\ \bibnamefont
  {Hou}}, \bibinfo {author} {\bibfnamefont {Z.}~\bibnamefont {Zhou}}, \bibinfo
  {author} {\bibfnamefont {X.}~\bibnamefont {Wang}}, \bibinfo {author}
  {\bibfnamefont {H.}~\bibnamefont {Guo}},\ and\ \bibinfo {author}
  {\bibfnamefont {C.-C.}\ \bibnamefont {Chien}},\ }\bibfield  {title} {\bibinfo
  {title} {Local geometry and quantum geometric tensor of mixed states},\
  }\href {https://doi.org/10.1103/PhysRevB.110.035144} {\bibfield  {journal}
  {\bibinfo  {journal} {Phys. Rev. B}\ }\textbf {\bibinfo {volume} {110}},\
  \bibinfo {pages} {035144} (\bibinfo {year} {2024})}\BibitemShut {NoStop}%
\bibitem [{\citenamefont {Zhou}\ \emph {et~al.}(2024)\citenamefont {Zhou},
  \citenamefont {Hou}, \citenamefont {Wang}, \citenamefont {Tang},
  \citenamefont {Guo},\ and\ \citenamefont {Chien}}]{zhouprb}%
  \BibitemOpen
  \bibfield  {author} {\bibinfo {author} {\bibfnamefont {Z.}~\bibnamefont
  {Zhou}}, \bibinfo {author} {\bibfnamefont {X.-Y.}\ \bibnamefont {Hou}},
  \bibinfo {author} {\bibfnamefont {X.}~\bibnamefont {Wang}}, \bibinfo {author}
  {\bibfnamefont {J.-C.}\ \bibnamefont {Tang}}, \bibinfo {author}
  {\bibfnamefont {H.}~\bibnamefont {Guo}},\ and\ \bibinfo {author}
  {\bibfnamefont {C.-C.}\ \bibnamefont {Chien}},\ }\bibfield  {title} {\bibinfo
  {title} {Sj\"{o}qvist quantum geometric tensor of finite-temperature mixed
  states},\ }\href {https://doi.org/10.1103/PhysRevB.110.035404} {\bibfield
  {journal} {\bibinfo  {journal} {Phys. Rev. B}\ }\textbf {\bibinfo {volume}
  {110}},\ \bibinfo {pages} {035404} (\bibinfo {year} {2024})}\BibitemShut
  {NoStop}%
\bibitem [{\citenamefont {Wang}\ \emph {et~al.}()\citenamefont {Wang},
  \citenamefont {Lu}, \citenamefont {Liu}, \citenamefont {Ding},\ and\
  \citenamefont {Fu}}]{xgwang}%
  \BibitemOpen
  \bibfield  {author} {\bibinfo {author} {\bibfnamefont {X.}~\bibnamefont
  {Wang}}, \bibinfo {author} {\bibfnamefont {X.-M.}\ \bibnamefont {Lu}},
  \bibinfo {author} {\bibfnamefont {J.}~\bibnamefont {Liu}}, \bibinfo {author}
  {\bibfnamefont {W.}~\bibnamefont {Ding}},\ and\ \bibinfo {author}
  {\bibfnamefont {L.}~\bibnamefont {Fu}},\ }\bibfield  {title} {\bibinfo
  {title} {Double {{Wilczek-Zee}} connection and mixed-state quantum geometric
  tensor},\ }\href@noop {} {\bibinfo  {journal} {Chinese Physics B, accepted}\
  }\BibitemShut {NoStop}%
\end{thebibliography}%
\end{document}